\documentclass[preprint,showpacs,preprintnumbers,unsortedaddress,amsmath,amssymb]{revtex4}
\usepackage{graphicx}
\newcommand{\md}{{\mathrm d}}
\newcommand{\lP}{\ell_{\mathrm P}}
\newcommand{\be}{\begin{equation}}
\newcommand{\ee}{\end{equation}}

\begin{document}

\preprint{AEI--2004--065}
\preprint{IGPG-04/8-2}
\title{Coordinate Time dependence in Quantum Gravity}
\author{Martin Bojowald\footnote{e-mail address: {\tt mabo@aei.mpg.de}}}
\affiliation{Max-Planck-Institut f\"ur Gravitationsphysik, Albert-Einstein-Institut,
Am M\"uhlenberg 1, D-14476 Potsdam, Germany}

\author{Parampreet Singh\footnote{e-mail address: {\tt singh@gravity.psu.edu}}}
\affiliation{Institute for Gravitational Physics and Geometry, Pennsylvania State University, 
University Park, PA 16802, USA}

\author{Aureliano Skirzewski\footnote{e-mail address: {\tt skirz@aei-potsdam.mpg.de}}}
\affiliation{Max-Planck-Institut f\"ur Gravitationsphysik, Albert-Einstein-Institut,
Am M\"uhlenberg 1, D-14476 Potsdam, Germany}
 
\begin{abstract}
 The intuitive classical space-time picture breaks down in quantum
 gravity, which makes a comparison and the development of
 semiclassical techniques quite complicated. 
 Using ingredients
 of the group averaging method to solve constraints one can
 nevertheless introduce a classical coordinate time into the quantum
 theory, and use it to investigate the way a semiclassical continuous
 description emerges from discrete quantum evolution. Applying this
 technique to test effective classical equations of loop cosmology and
 their implications for inflation and bounces, we show that the
 effective semiclassical theory is in good agreement with the quantum
 description even at short scales.
\end{abstract}


\maketitle

\section{Introduction}

Current knowledge of the quantum structure of space-time suggests a
picture very different from the smooth classical one. Space and time
are expected to be fundamentally discrete such that both change only
in steps. Still, a transition between both pictures must be possible
in order to understand the emergence of a classical world on large
scales from the fundamental quantum world. Also for practical purposes
such a transition is helpful in a semiclassical approximation.
A technically and conceptually important question is where the
classical picture starts to make sense or, when going to smaller and
smaller scales, where it breaks down. The change of scale can happen
either computationally, i.e.\ by looking at smaller and smaller scales
in a coarse-grained approximation which can then be used, for
instance, to understand the breaking or deformation of classical
symmetries, or dynamically during the expansion or contraction of a
universe or the collapse of matter to a black hole.

One application is the behavior of universes which on larger scales
has been studied from the point of view of loop quantum cosmology
\cite{LoopCosRev,ICGC} by using effective classical equations
\cite{Inflation, Closed,Spin,EffHam}. While the fundamental
description is quantum, governed by a difference equation for the wave
function \cite{cosmoIV,IsoCosmo,HomCosmo}, effective classical
equations show the diverse cosmological effects more easily. The idea
of using effective classical equations is that in semiclassical
regimes they describe the position of wave packets solving the
difference equation. They simplify the analysis considerably even in
isotropic models and can be expected to do so even more in
inhomogeneous models or the full theory. In light of the previous
discussion an open question is where exactly an effective classical
equation makes sense as a good approximation to the behavior of the
difference equation, and where additional correction terms have to be
taken into account.

This question can be answered by a direct comparison of effective
semiclassical descriptions, given by ordinary differential equations,
with the underlying discrete quantum evolution governed by difference
equations (these difference equations may even be partial depending on
the number of matter fields). However, since an ordinary differential
equation is quite different from a discrete difference equation, their
solutions can not be compared directly.  For such a purpose we first
have to extract appropriate data from solutions of the difference
equation, usually by taking expectation values, which we then compare
to the classical theory or one with further corrections.  (At this
point one has to distinguish between ambiguities resulting from
choosing how to extract the semiclassical data, and outright
deviations from the purely classical behavior. How this can be
disentangled will be discussed later.) In this way one can see if new
effects arise or in which range one can trust an effective classical
equation with or without certain correction terms.

Comparing an effective theory with the underlying quantum theory is of
great relevance to various issues in loop cosmology. For instance in
order to determine the starting point of inflation
\cite{Inflation,InflationWMAP,Robust,GenericInfl}, observable
signatures in cosmic microwave background \cite{InflationWMAP}, the validity of effective
classical bounce pictures
\cite{BounceClosed,Oscill,BounceQualitative,Cyclic,GenericBounce},
general aspects of the approach to a classical singularity
\cite{NonChaos,ChaosLQC} or evolution of perturbations.  If the
effective theory requires further corrections from the quantum theory
which can be verified by their comparison, then these correction terms
in principle can leave an observational signature which can be used to
verify loop quantum cosmology.

A particularly striking difference between the classical and the
quantum theory is the issue of time. A common understanding which
works in both cases is that of relational time, where time is not an
external, absolute parameter but encoded in the relative change
between different degrees of freedom
\cite{BergmannTime,KucharTime,RovelliTime}. However, this concept is
difficult to use explicitly, and so classically one employs the
space-time picture where time is just a gauge coordinate. Thus, this
time coordinate has no invariant physical meaning, but nevertheless
provides a helpful intuitive understanding of a given gravitational
system. From the Hamiltonian point of view, this time coordinate is
the gauge parameter for orbits generated by the Hamiltonian
constraint. In this way, coordinate time is related to the second
effect of a first class classical constraint, namely that of
generating a gauge transformation in addition to restricting fields to
the constraint surface.

In a canonical quantum theory the situation is different because in
the Dirac procedure to solve first class constraints there is only one
step by requiring physical states to be annihilated by the quantum
constraints, which then are automatically gauge invariant. In systems
with a Hamiltonian constraint, physical states are thus timeless,
which has led to the name `frozen formalism.' As we will discuss
below, the two steps familiar from the classical procedure can easily
be disentangled also in a quantum procedure to solve the constraints,
in particular if the technique of group averaging \cite{Refined} is
used. 
Physical states can then be represented in an evolving manner,
depending on the gauge choice via the lapse function. 
We emphasize that our notion of evolution in time as used here is with
respect to coordinate time 
and not physical time.  We call this coordinate time-dependent
family of states a `state-time' in order to indicate that classical
space in a space-time has been replaced by a quantum state, while time
remains classical.  At a given time parameter of the state-time the
constraints will not be satisfied, but the whole state-time represents
a physical state, which can be reconstructed by integrating over time,
in a well-defined way.

As a practical application we develop a scheme to decide the domain of
validity of an effective semiclassical description and whether in some
regimes it requires further correction terms.
For that, we compare the expectation value of,
e.g., the volume in a given 
state-time
with the volume obtained from the classical theory (or with an
effective theory including further corrections). The correction terms
can be derived by computing the expectation value of the Hamiltonian
constraint operator in a coherent state and expanding around the
classical expression \cite{Bohr,Perturb}. There are diverse sources of
deviations in the case of non-linear constraints which can be studied
analytically or numerically: First, there are the usual Ehrenfest
statements about the relation between classical and quantum equations
of motion for expectation values. In addition there are choices
related to choosing an initial semiclassical state, and the way the
constraints are violated at fixed coordinate times.  Finally and most
importantly, there are genuine quantum gravity corrections whose
implications can give rise to new physical effects. The latter imply
the modifications we want to include in effective classical equations.

As we will see, it is possible to disentangle these effects at least
qualitatively. The resulting modifications to classical behavior can
then be compared to known analytical results, as done here for
inflationary behavior and bounces, or be used to suggest new effects.
In the next section we recall the group averaging procedure which will
be followed by a brief discussion of the issue of introducing
coordinate time in this context. In Sec.\ IV we will present a way to
implement this idea numerically and study examples of universes with a
cosmological constant or dust as matter.  We will show that, both for
inflation and bounces, the effective semiclassical theory which
incorporates modified geometrical densities is in good agreement with
discrete quantum evolution till very small scales. This proves that
effects derived from the effective semiclassical theory capture to a
large extent the true nature of quantum spacetime.  Here we focus on
describing the way our technique can prove useful in testing a
semiclassical theory, leaving a more detailed study for future work.

\section{Group averaging}

We will discuss here only systems with a single constraint which we
mostly think of as a Hamiltonian constraint generating coordinate
time. In an isotropic cosmological model the constraint itself will be
the Friedmann equation, and the evolution equation it generates is the
Raychaudhuri equation plus equations for matter fields such as the
Klein--Gordon equation for a scalar. After quantizing we obtain the
constraint operator which annihilates physical states, and all
evolution would have to be extracted from physical states in a
relational way.

A simple method to implement a constraint on quantum states is as
follows. Consider the action of the constraint operator $\hat C$ on a
given non-physical state interpreted to promote a change in gauge
parametrized by a real parameter $\lambda$
\begin{equation} \hat C\left|
\varphi_\lambda\right>=i\frac{\md}{\md\lambda}\left|
\varphi_\lambda\right>\,.
\end{equation} 
Obviously $\left|
\varphi_\lambda\right>$ is not annihilated by the operator $\hat{C}$
unless $|\varphi_{\lambda}\rangle$ is independent of $\lambda$, i.e.\
gauge invariant. To achieve $\lambda$-independence we
can average the state $\left|
\varphi\right>\equiv
\int \md\lambda\left| \varphi_\lambda\right>$
such that
\begin{equation}
 \hat C \left| \varphi\right> = \int \md\lambda \, \hat C\left|
 \varphi_\lambda\right>=i\int \md\lambda \, \frac{\md}{\md\lambda}\left|
 \varphi_\lambda\right>
\end{equation}
which in the case of compact symmetry orbits or suitable boundary
conditions in the non-compact case vanishes identically. In
conclusion, we can get a solution to our constraint by averaging the
nonphysical state over the symmetry group: $\int \md\lambda\left|
\varphi_\lambda\right>$. This is in fact not new because the solution
to our equation for a state $\left| \varphi_{\lambda}\right>$ is
given by
\begin{equation}
 \left| \varphi_\lambda\right>=e^{-i\lambda \hat{C}}
 \left| \varphi_0\right>
\end{equation}
and the invariant state is 
\begin{equation}
\int \md\lambda \, e^{-i\lambda \hat{C}}\left| \varphi\right>
\end{equation}
which is nothing else than the group averaging map \cite{Refined} into
the physical Hilbert space.

As examples we consider different cases in which the method works with
different success.  The simplest case would be to consider the
constraint $\hat P_\theta \left|\varphi\right>=0$ which imposes
rotation invariance on a two dimensional system with angular coordinate
$\theta$. Then, by following the prescription presented above we
replace the equation
\begin{equation}
 \left<\theta\right|\hat P_\theta
 \left|\varphi\right>=i \, \partial_\theta\varphi(\theta)=0
\end{equation}
with a Schr\"odinger like equation. The constraint then acts on
non-physical states by
\begin{equation}
 i \, \partial_\theta\varphi_\lambda(\theta)=
 i \, \partial_\lambda\varphi_\lambda(\theta)
\end{equation}
with general solution
\begin{equation}
 \varphi_\lambda(\theta)=\varphi(\lambda+\theta)\,.
\end{equation}
Physical states are thus given by 
\begin{equation}
 \varphi=\int_0^{2\pi} \md\lambda \, \varphi(\lambda+\theta)
 =\int_0^{2\pi} \md\lambda \, \varphi(\lambda)
\end{equation}
where any $\theta$-dependence is removed.

We can also consider the non compact case by using the translation
generator $\hat P_x$ with exactly the same calculations (except that
allowed functions $\varphi$ have to be restricted to a suitable set
for the $\lambda$-integration to exist, similarly to selecting an
appropriate subspace of the kinematical Hilbert space for group
averaging).  For a more general example let us consider the operator
$\hat C = a
\hat X\hat P+b$ to follow the strategy above, i.e.\ use the
operator to generate transformations of an arbitrary state, and then
solve the resulting partial differential equation. In this case the
equation becomes
\begin{equation}
 \left(a \, x \, i\partial_x+b\right)\varphi_\lambda(x)=
 i\partial_\lambda\varphi_\lambda(x)
\end{equation}
with solution
\begin{equation}
 \varphi_\lambda(x)=
 e^{\frac{ib}{2}(\frac{1}{a}\ln(x)-\lambda)}f(\ln(x)/a+\lambda).
\end{equation}
Then, in order to realize the $\lambda$ integration we Fourier
transform $f(\frac{1}{a}\ln(x)+\lambda)$ which is possible only for $a$
real: $f(u)=(2\pi)^{-1}\int_{-\infty}^{\infty}\md\omega e^{-i\omega
u}\tilde{f}(\omega)$. Moreover, we commute the $\lambda$- and
$\omega$-integrations,
\begin{equation}
 \int_{-\infty}^{\infty}\md\lambda\varphi_\lambda(x)=
 e^{i\frac{b}{2}\frac{1}{a}\ln(x)}
 \int_{-\infty}^{\infty}\frac{\md\omega}{2\pi} \, e^{-i\omega\frac{1}{a}\ln(x)}
 \tilde{f}(\omega) \int_{-\infty}^{\infty}\md\lambda \,
 e^{-i(\omega+\frac{b}{2})\lambda}\,.
\end{equation}
Here, the integral in $\lambda$ would diverge for arbitrary complex
$b$ and we get a solution for the constraint only if it is real. After
integrating over $\lambda$ and $\omega$ we obtain
$\varphi(x)=ce^{i\frac{b}{a}\ln(x)}$ which can be checked to satisfy
our constraint. If $b\not\in{\mathbb R}$, the final integration
diverges which means that we are not allowed to commute the
integrations. Moreover, in this case we would have to choose
appropriate fall-off conditions for $f$.

Whether or not we are using a self-adjoint constraint operator has
significance for the physical inner product, which we are not
considering here. Still, adjointness properties also play a role at
the level of solving the constraint, as the following example given by
$\hat{C}=\partial/\partial x$ demonstrates. Now we have to solve the
equation
\[
 \partial_x
\varphi_{\lambda}(x)=i \, \partial_{\lambda}\varphi_{\lambda}(x)
\]
which is done by any arbitrary function
$\varphi_{\lambda}(x)=\varphi(\lambda+ix)$ depending only on
$\lambda+ix$. Integrating over $\lambda$ to compute a physical state
is now done along a line shifted vertically by the amount $x$ in the
complex plane, where we interpret solutions $\varphi_{\lambda}(x)$ as
holomorphic functions on the complex plane with coordinate
$z=\lambda+ix$. The result will be independent of $x$ only if
$\varphi$ satisfies appropriate fall-off conditions and does not have
residues. Both properties together cannot be satisfied for non-zero
functions since, owing to Liouville's Theorem any bounded entire
function is constant. Thus, any function for which the
$\lambda$-integrations exist must have poles on the complex plane such
that the averaging procedure does not lead to constant functions of
$x$, as expected for this constraint, but only to locally constant
ones.

While the averaging can be defined even for non-selfadjoint
constraints, from the numerical point of view self-adjointness of the
constraint operator is essential. Non-real eigenvalues would imply
exponentially growing modes in solutions to the differential equation
which lead to numerical instabilities.

\section{Coordinate time}

Quantum gravity in the perspective of canonical quantization arises as
a constrained system which gives rise to the well known
Wheeler--DeWitt equation instead of a Schr\"odinger like
equation. Nowadays canonical quantum gravity is usually formulated
with Ashtekar variables and it is possible to realize the
constraint algebra on a well-defined kinematical Hilbert space
\cite{ALMMT,QSDI,QSDII,QSDIII,Master}. The Gauss and diffeomorphism
constraints can be solved by group averaging \cite{ALMMT}, but
discussions about the correct Hamiltonian constraint are not settled
yet \cite{QSDI,LM:Vertsm,Consist,Master}. Also the issue of the
physical inner product and how to use the solution space to the
Hamiltonian constraint are almost completely open in the full
theory. After reducing to symmetric models \cite{SymmRed}, the
Hamiltonian constraint simplifies and can often be treated
explicitly. Even in the simplest cosmological models the physical
inner product is not yet understood, but since the spatial volume can
be used as internal time in a cosmological situation the problem of
time does not play a role. In all these cases there is a Hamiltonian
constraint whose gauge parameter classically corresponds to coordinate
time, and it is our aim to discuss how such a parameter can appear
with this interpretation in quantum theory.
We emphasize that, as a general problem, this is much simpler than the
problem of time where time is understood as a physical parameter valid
for the full quantum theory. In contrast, we are interested in
formulating the quantum theory in a parameterized way, with a new
non-physical coordinate time parameter. Moreover, this parameter is
expected and intended to make sense only in semiclassical regimes.
This will be discussed in more detail in the Conclusions making use of
what we learned in the examples.

As discussed before, the group averaging procedure to solve
constraints can be split into two steps which roughly correspond to
the two classical steps of restricting to the constraint surface and
factoring out by the gauge orbits. The correspondence is not perfect,
however, since one single member $\varphi_{\lambda_0}$ of a family
$\{\varphi_{\lambda}\}_{\lambda\in{\mathbb R}}$ of states exhibiting
the gauge parameter,
which we call {\em state-time} in the case of a gravitational system,
is not a solution to the quantum constraint, while a classical gauge
orbit can completely lie within the constraint surface. Even in a
semiclassical regime this would lead to deviations between the
classical and quantum behavior since the constraints are always
violated when the gauge parameter is exhibited. Still, as an
approximation and a heuristic tool the gauge dependent family of
states can be very useful. In particular for the Hamiltonian
constraint of a gravitational system this allows us to describe the
quantum dynamics approximately (in the sense that the constraint is
not imposed exactly for otherwise the state-time would have to be
time-independent)
with reference to a coordinate time,
which justifies the use of effective classical equations of motion.

Taking into account possible choices of lapse functions, which after
quantizing can be operators if $N$ depends on $t$ via kinematical
degrees of freedom (such as $N(t)=a(t)$ which is used when
transforming from proper time to conformal time in an isotropic
model), we arrive at
\begin{equation} \label{Ham}
\hat N\hat H\left| \Psi_t\right>=i\frac{\md}{\md t}\left| \Psi_t\right>
\end{equation}
as the evolution specified by the Hamiltonian constraint
operator. 

The condition that the Hamiltonian constraint has to annihilate
physical states emphasizes the fact that physics does not depend on
(coordinate) time. Correspondingly, Eq.\ (\ref{Ham}), which is
directly related to a choice of time through a Schr\"odinger equation
on non physical states, and the state-time $|\Psi_t\rangle$ solving it
are in fact not unique because we can fix the gauge freedom in
different ways by making different choices of the quantized version
$\hat N$ of the lapse function.

If we were interested in completing the group averaging we would have
to integrate the state-time over $t$ in order to arrive at a physical
state. However, just as the classical space-time picture, which we are
interested in here, arises only when gauge orbits parameterized by
coordinate time $t$ are not factored out, we have to refrain from
performing this final step and instead work only with the state-time.
In principle, with much effort one can always remove all gauge
dependence, classically by factoring out gauge and in quantum theory
by integrating over the gauge group, but physical intuition is best
developed in a (coordinate) time dependent picture. Nevertheless, we
think that the discussion of group averaging justifies our use of the
state-time and the following applications.

\section{Applications}

For semiclassical physics it is very convenient to have an explicit
coordinate time parameter since classical intuition is based on the
space-time picture. In principle it is also possible to work with
internal times both at the quantum and classical level, but it comes
with much more technical effort. In the case of loop quantum cosmology,
in fact, most recent phenomenological applications
\cite{Inflation,InflationWMAP,BounceClosed,NonChaos,ChaosLQC,Robust,Oscill,BounceQualitative,Cyclic,GenericInfl,EffHam,GenericBounce}
are based on effective classical equations \cite{Inflation,Closed}
which are differential equations in coordinate time and implement the
main non-perturbative quantum effect \cite{InvScale,Ambig} in matter
Hamiltonians \cite{QSDV}. Direct studies of the underlying difference
equations, on the other hand, are more complicated
\cite{Scalar,ScalarLorentz,FundamentalDisc,GenFunc,ClosedExp}.

These equations show the main qualitative effects that have to be
expected from the loop quantization at a more intuitive level. For
more quantitative applications, however, it is important to see to
which degree they provide an approximation
(in the sense discussed in the Introduction)
to the quantum behavior and whether in some regimes additional
correction terms have to be taken into account. 
One way of evaluating these correction terms for instance is to
compute the expectation value of the constraint operator in a coherent
state \cite{Bohr,Perturb}. This results in the classical constraint
together with correction terms, and thus corrected equations of motion
for, e.g., the scale factor. The present paper suggests an alternative
procedure, which is one step closer to the quantum theory. Using the
quantum coordinate time picture, we can evolve the state first within
the quantum theory, and then compute the expectation value of, e.g.,
the volume operator which also gives us the time dependence of the
scale factor. Both procedures are approximations to the quantum
dynamics: In the first case one uses kinematical coherent states to
compute the expectation value, while in the second case coordinate
time is introduced which, as explained above, is not exact in quantum
theory. In both cases, however, the Hamiltonian constraint is used and
implemented at least partially: The first case imposes a classical
constraint with quantum corrections, while in the second the quantum
constraint is used to evolve states, which then could be averaged if
we are interested in the physical state.

In a sense, the procedures differ by a commutation of evolving and
translating from quantum to classical behavior (by taking expectation
values). We either take the expectation value of the constraint first
(in a kinematical coherent state) and then determine the evolution in
a classical manner, or we first evolve a kinematical state with
quantum operators and then determine classical quantities from
expectation values in the resulting, time-dependent states. It is not
guaranteed that the two different steps in fact commute, which leads
to differences between the two procedures. In simple models, however,
both ways of determining the dynamics can be implemented at least
numerically and then compared with each other. We will demonstrate
this in what follows, leaving a more detailed investigation with
precise statements of ranges of applicability and of the necessity of
additional correction terms for future work.

\subsection{Numerical implementation}
\label{NumImp}

In isotropic loop quantum cosmology for a flat model, the Hamiltonian
constraint operator is given by \cite{IsoCosmo,Bohr}
\begin{eqnarray*}
 (\hat{H}_0\tilde\psi)_{\mu} &=& (V_{\mu+5}- V_{\mu+3})\tilde\psi_{\mu+4} -
 2( V_{\mu+1}- V_{\mu-1})\tilde\psi_{\mu} \\
 &&+ ( V_{\mu-3}- V_{\mu-5})\tilde\psi_{\mu-4}
 +\tfrac{8}{3}\pi G\gamma^3\ell_{\rm P}^2\,\,
 {\rm sgn}(\mu)\hat{H}_{\rm matter}(\mu) \tilde\psi_{\mu}
\end{eqnarray*}
acting on a wave function $\tilde{\psi}\colon{\mathbb R}\to{\mathbb
C}$ supported on eigenspace of the triad operator $\hat p$. The
coefficients are given in terms of the volume eigenvalues
$V_{\mu}=(\frac{1}{6}\gamma\ell_{\rm P}^2)^{3/2}|\mu|^{3/2}$, with the
Barbero--Immirzi parameter $\gamma=0.238$ \cite{Gamma,Gamma2} as it
follows from calculations of black hole entropy
\cite{ABCK:LoopEntro,IHEntro}, and $\hat{H}_{\rm matter}(\mu)$ is the
matter Hamiltonian which we will choose later.

Since only labels with distances of four apart are involved, we introduce
$4m:=\mu$ with integer $m$ (see also \cite{Velhinho} for this
restriction) and write the operator as
\begin{eqnarray*}
 (\hat{H}_0\psi)_m &=& ( v_{m+5/4}- v_{m+3/4})\psi_{m+1} -
 2(v_{m+1/4}- v_{m-1/4})\psi_m \\
 &&+( v_{m-3/4}-v_{m-5/4})\psi_{m-1}
 +\tfrac{8}{3}\pi G\gamma^3\ell_{\rm P}^2\,\,
 {\rm sgn}(m)\hat{H}_{\rm matter}(4m) \psi_m
\end{eqnarray*}
with $\psi_m:=\tilde\psi_{4m}$ and $v_m:= V_{4m}$. This operator will
have to be symmetrized to $\hat{H}:=\frac{1}{2}
(\hat{H}_0+\hat{H}_0^{\dagger})$ for 
the numerical implementation of coordinate time to be stable.
(Alternatively, we can choose a lapse function
$N(t)=a(t)$, corresponding to conformal time, and quantize such that
$\hat{N}$ has eigenvalues proportional to $V_{\mu+1}-V_{\mu-1}$. The
resulting operator $\hat{N}\hat{H}_0$ would be symmetric without
reordering $\hat{H}_0$, but since the lapse function and its
quantization vanish for $a=0$, the quantum behavior around $\mu=0$
will be problematic.)  


For numerical purposes we need to
restrict the operator to finite lattices ${\cal
L}_{m_c,N}:=\{n\in{\mathbb Z}:m_c-N/2< m \leq m_c+N/2\}$ of size $N$
and centered at $m_c$, such that it will be represented by a
tridiagonal $N\times N$-matrix $H_{ij}$ with $H_{ij}=0$ for $j>i+1$ or
$i>j+1$,
\begin{equation}
 H_{ii}=-2 (v_{m_c-N/2+i+1/4}-v_{m_c-N/2+i-1/4})+ \tfrac{8}{3}\pi
 G\gamma^3\ell_{\rm P}^2\: {\rm sgn}(m_c-\tfrac{1}{2}N+i) \hat{H}_{\rm
 matter}(m_c-\tfrac{1}{2}N+i) \label{diffeq1}
\end{equation}
and
\begin{equation}
 H_{i,i+1}=H_{i+1,i}= \tfrac{1}{2}(v_{m_c-N/2+i+5/4}- v_{m_c-N/2+i+3/4}+
v_{m_c-N/2+i+1/4}- v_{m_c-N/2+i-1/4})\,. \label{diffeq2}
\end{equation}
The above splitting of Hamiltonain matrix into diagonal and off-diagonal parts is
done purely for numerical convenience.
We now introduce the coordinate time parameter $t$ and solve
\begin{equation} \label{diffHam}
 H\cdot\psi_t=i \, \frac{\md}{\md t}\psi_t
\end{equation}
numerically. The solution to the differential equation
(\ref{diffHam}) is then given by
\[
 \psi_t=\exp(-itH)\cdot\psi_{\rm in}
\]
with an initial state $\psi_{\rm in}\in{\mathbb C}^N$. 
The function $\psi_t$ can then be used to compute time dependent
expectation values.

\subsection{Examples}

As examples of our technique, we would now investigate various cases
of matter Hamiltonian and compare the effective semiclassical theory
with the evolution determined by quantum difference equations. We
first consider the case where matter Hamiltonian is just a
cosmological constant and the scale factor is the sole degree of
freedom.  Unlike with an internal time evolution, for which no other
degree of freedom besides internal time $a$ would be left, this still
allows us to have a non-trivial evolution of $a$ in coordinate time
$t$. This example is then followed by examples of inflation and bounce
with inclusion of matter (dust) in the analysis which signifies the
role played by effective densities in a good semiclassical
description.

The total Hamiltonian is given by
\be
\frac{3}{2} \, a \dot a^2 - 8 \pi G \, H_{\rm matter} = 0
\ee
which leads to the Friedmann equation
\be
\frac{\dot{a}^2}{a^2} = \frac{16\pi G}{3} \, \rho(a) ~. \label{effth}
\ee
In our first example we take $\rho(a)=\Lambda$ and later we will
consider the classical dust density $\rho(a)=M/a^3$.
%
In general,
there are two qualitatively different kinds of modifications which bring
the behavior of effective cosmological equations closer to that of
the quantum theory, which we will study in what follows. First, by
non-perturbative effects the geometrical density $a^{-3}$ in the matter
Hamiltonian density is replaced by a function which is finite and does not
diverge at small $a$. The precise form of the function will be
discussed below.  Secondly, there are perturbative corrections in the
gravitational part $\dot{a}^2/a^2$ of the constraint which appear as
additional terms on the left hand side of the Friedmann equation. We
will mainly discuss one such term, which includes effects of the
spread of a wave packet in the effective classical framework.

\subsubsection{Effective classical behavior: Cosmological constant}

In order to compare the evolution given by an effective theory and
the quantum difference equation we first have to choose an initial state
$\psi_{\rm in}$.
This state should be peaked
on a specified classical volume, together with an extrinsic curvature
which follows from the volume and the classical Hamiltonian constraint
to which we compare the quantum evolution. In principle, one can use any
suitable state as an initial state, however for simplicity we will use
a Gaussian
\[
 \psi_{{\rm in},k}=\exp\left(-\frac{(k-N/2)^2}{4\sigma^2}+2i(k-N/2)c_0\right)
\]
in what follows. Here the isotropic connection component $c_0$ is
related to extrinsic curvature by $c=-\gamma\dot{a}/2$. It is determined
from the peak scale factor $a_0=\sqrt{\frac{2}{3}\gamma\lP^2|m_c|}$
\footnote{The loop quantization of the symmetry reduced cosmological
  model from the full theory is done in such a way that the scale
  factor as used here (which is related to the cubic root of the
  volume of the fiducial cell necessary to define symplectic
  structure) is invariant to its conventional rescaling freedom in
  classical cosmology. For details see Ref.\ \cite{Bohr}.}  by the
constraint $-6c_0^2a_0+8\pi G\gamma^2 H_{\rm matter}(a_0)=0$. (At this
point the classical equations to be compared with enter. If correction
terms are included, as will be done later, the value of $c_0$ changes
and so does the quantum evolution of the new initial state.)
Nevertheless, the explicit choice has an influence on the evolution
such as the degree of spreading of the wave packet. Numerically, a
choice leading to less spreading will allow the evolution to be
reliable for a longer period of time since the boundary values will
remain small longer and finite size effects will set in later.

For our purpose, comparing the quantum evolution with classical
equations, the choice of initial wave packet is not that important
since it is sufficient to know the evolution for limited amounts of
time only. We will not be able to compare whole solutions as functions
of time in this way, but since the classical equations and also the
corrected ones are local in time we can study  deviations. One
can then decide which correction terms are needed to describe the
local change in volume following from the quantum evolution.

\begin{figure}
 \includegraphics[width=14cm,keepaspectratio]{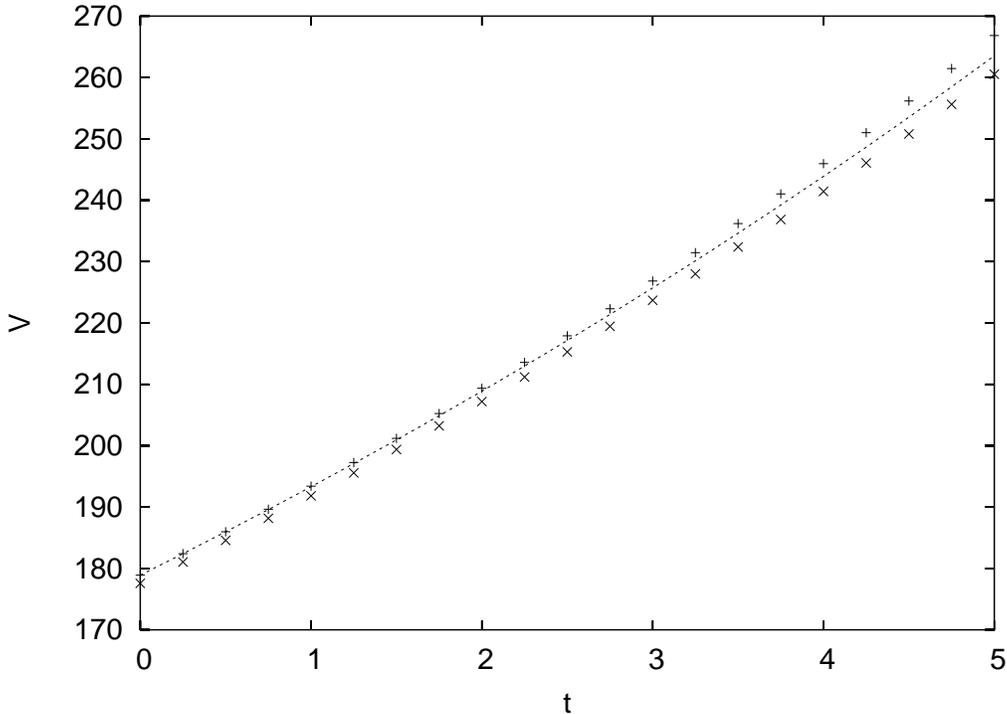}
 \caption{Expectation value of the volume ($+$) compared to the
effective  solution $V(t)$ (dashed) and the volume $\langle
 \hat{a}\rangle^3(t)$ computed from the expectation value of the scale
 factor ($\times$) for the case of cosmological constant. 
The initial peak of the coherent state is chosen around 
$m_c=200$ with $\sigma=20$ and $N=500$. We take $\Lambda=10^{-3}$. \label{LambdaSmall}}
\end{figure}

A quantitative statement about the deviations is complicated by the
fact that the quantum evolution is not uniquely related to a classical
expression. One can take expectation values and compare with the
classical functions, but due to the spread of the probability
distribution the result depends on whether we take, e.g., the
expectation value
\begin{equation}
 \langle\hat{V}\rangle(t)=||\psi_t||^{-2} \sum_{k=1}^N
 v_{m_c-N/2+k}|\psi_{t,k}|^2
\end{equation}
of the volume operator, or that of the scale factor operator
$\hat{a}=\hat{V}^{1/3}$ and the cubic power afterwards.  

\begin{figure}[tbh!]
 \includegraphics[width=16cm,keepaspectratio]{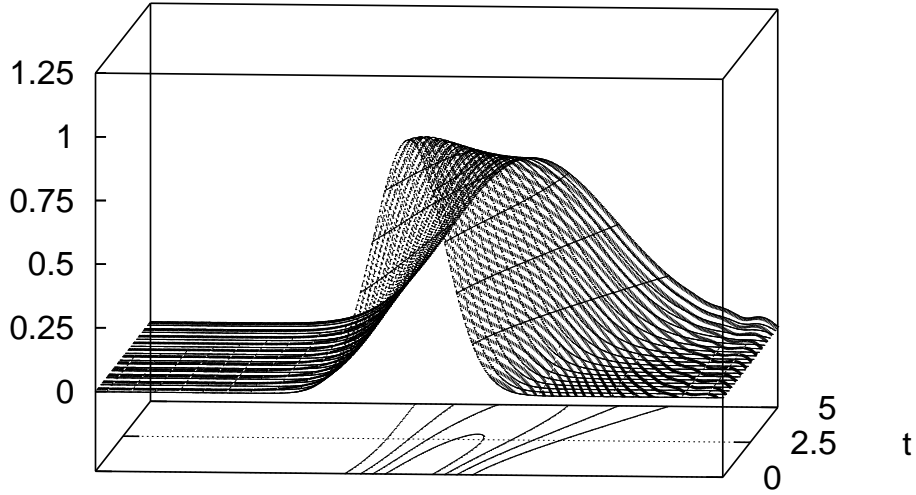}
 \caption{A spreading wave packet for $\Lambda=10^{-3}$ as in
 Fig.~\ref{LambdaSmall}. At the back of the right hand side one can
 see small oscillations building up when the wave reaches the
 boundary. The vertical axis shows the magnitude of the normalized wave
 function, which starts as a Gaussian at the front and evolves to the
 back. \label{LambdaWave}}
\end{figure}

Using these ingredients we compute the volume for the quantum
equation (\ref{diffHam}) and the classical theory (eq.\ref{effth}).
Classically, we have $V=a^3$, but this relation will certainly not be
satisfied for the expectation values of $\hat{V}$ and $\hat{a}$ once
the spread of the wave function becomes large. In
Fig.~\ref{LambdaSmall}, we have shown the behavior of expectation
values of the volume operator $\hat V$ ($+$) and that evaluated from
$\hat a$ ($\times$). These are compared with the volume calculated
from the classical theory (dashed curve). As it is clear, the
classical theory gives a very good approximation to the underlying
evolution from the difference equation. However, due to spread of the
wavepacket some discrepancy occurs for late times. This can also be
noticed in the evolution of the wave packet which is shown in
Fig.~\ref{LambdaWave}. Comparing the two different ways of computing
volume expectation values shows that the discrepancies do not result
from new physical effects but only reflect the ambiguous way of
relating expectation values to classical behavior.

In general, departures between expectation values of the volume
operator and classical values can have several reasons.  Besides
approximations used in the method to solve the constraint, there are
quantum effects which we are most interested in here. They can be
divided into two classes, the first one arising from small-scale or
high-curvature effects in quantum operators, the second one coming
from the fact that we have an evolving wave packet rather than a sharp
classical point. Both effects can be included into effective classical
equations, but in the second case it is not always clear if
modifications to the classical behavior are a consequence of having
chosen a bad initial state or a physical effect related to the
evolution of spread, asymmetry and other deformations in profile of
the probability distribution.

\begin{figure}[tbh!]
 \includegraphics[width=16cm,keepaspectratio]{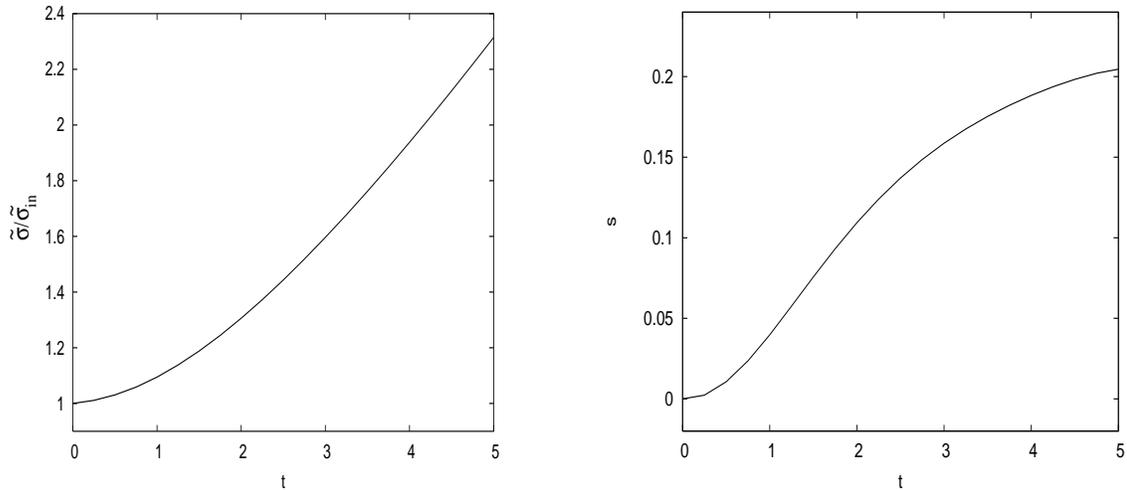}
 \caption{Growth of standard deviation ($\tilde\sigma/\tilde\sigma_{in}$) and skewness ($s$) for the cosmological constant
case studied in Fig. \ref{LambdaSmall}. The coherent state remains almost peaked over the classical value
due to small growth in skewness. \label{LambdaSmall_spread}}
\end{figure}

In order to quantify this, one can evaluate the skewness of the
wavefunction as it evolves. The skewness which is initially zero for
the Gaussian coherent state describes the asymmetry of the wave
packet and is given in terms of various expectation values of powers
of $\hat p=\hat{V}^{2/3}$ as
\begin{equation}
s = \frac{1}{\tilde \sigma^3} \, \bigg[\langle \hat p^3 \rangle - 3 \,
\langle \hat p \rangle \, \langle \hat p^2 \rangle + 2 \langle \hat p
\rangle^3 \bigg] \label{skew_eq}
\end{equation}
where $\tilde \sigma$ is the standard deviation $(\langle \hat p^2
\rangle - \langle \hat p \rangle^2)^{1/2}$ which initially is given by
$\tilde{\sigma}_{\rm in}=\frac{2}{3}\gamma\sigma$. (We use $\hat{p}$
in order to define skewness because it is the isotropic component of
the densitized triad, which is basic in loop quantum
gravity. Eigenvalues of $\hat{p}$ are proportional to $\mu$ or $m$
such that the skewness is computed for the variable in the wave
function.)  It is clear from Fig.~\ref{LambdaSmall_spread} that the
initial Gaussian wave packet gets skewed with time and deforms, but
the deformation remains small such that the main parameters
characterizing the wave packet in the cases studied here are only its
expectation value and spread.

The evolution of skewness with time reflects the fact that the
coherent state is no longer peaked at its classical value, which is
one cause of departure between quantum and classical curves.
Fig.~\ref{LambdaSmall_spread} shows the spread of the wavefunction
with respect to its initial value. A symmetric spreading would give an
evolved coherent state still peaking at classical values at late times
for the operator $\hat{a}^2$ corresponding to the discrete argument
$m$ of the wave function. The significance of small growth in skewness
for the time scale of interest is that the probability distribution of
the wavefunction remains peaked close to the classical volume.
Moreover, since the quantized volume and scale factor are given by
powers of $m$ different from one, they do not follow the classical
curve exactly, their difference increasing with increasing spread and
skewness.

This discussion shows that a comparison between quantum and
(effective) classical behavior cannot be done arbitrarily precisely
because of the unsharp nature of quantum wave packets. To determine
the level up to which a comparison is reliable we have different
techniques as illustrated in this case. The spread of the
wave packet and its deformation can be computed and plotted such that
a strong growth signals stronger departures. Similarly, plotting
expectation values of different powers shows a window in which the
classical curve has to be expected if there would be no quantum
modifications to the equations of motion. As we will see in the
following examples, effects from such quantum modifications are much
stronger such that they can easily be separated from simple spread
effects.

\subsubsection{Accelerated expansion}

In the previous subsection on cosmological constant we showed that
evolution extracted from the quantum difference equation by
introduction of coordinate time agrees well with the evolution
determined classically.  Now we will study the case of dust
Hamiltonian by introducing the modification to geometrical density in
the Friedmann equation (\ref{effth}).  This modification to
geometrical density is a novel prediction of loop quantum cosmology at
short scales and has led to various interesting applications
\cite{Inflation,Closed,EffHam,
  Spin,cosmoIV,IsoCosmo,HomCosmo,InflationWMAP,Robust,GenericInfl,BounceClosed,Cyclic,
  Oscill,BounceQualitative,GenericBounce,NonChaos,ChaosLQC}.
The main new ingredient is then given
by a non-perturbative modification to the classically diverging density
$a^{-3}$ (which has been derived in \cite{Ambig} and is based on
expressions of the full theory \cite{QSDV}) relevant at
small volumes. If a matter Hamiltonian has a density term 
then such modifications would enter dynamics and change the
behavior at small scale factors. For example, the Hamiltonian for a massive
scalar field $\phi$ is given by
\be
H_{\phi}(a)=\frac{1}{2} \,a^{-3} \, p_{\phi}^2 \, + \, a^3 \, V(\phi)
\ee
with momentum $p_{\phi}$ and potential $V(\phi)$. This Hamiltonian 
in loop quantum cosmology is modified to
\be
H_{\phi}(a)=\frac{1}{2} \,d_{j,l}(a) \, p_{\phi}^2 \, + \, a^3 \, V(\phi)
\ee
where $d_{j,l}$ is the modified geometrical density given by
\begin{equation}
d_{j,l}(a)= a^{-3}D_l(3a^2/\gamma j\lP^2)
\end{equation}
where
\begin{eqnarray}
D_l(q) &=&q^{3/2} \left\{\frac{3}{2l}\left(\frac{1}{l+2}
\left[(q+1)^{l+2}-|q-1|^{l+2}\right]\right.\right. \nonumber\\
&& - \left.\left.\frac{q}{l + 1}
\left[(q+1)^{l+1}-{\rm sgn}(q-1)
|q-1|^{l+1}\right]\right)\right\}^{3/(2-2l)}
\end{eqnarray}
and $j$ is a quantization ambiguity parameter (a half-integer)
\cite{Ambig}. There is another ambiguity parameter $0<l<1$
\cite{ICGC}, which is more restricted by full loop quantum gravity and
is usually taken as $l=3/4$. For very small $a\ll \sqrt{j}\lP$,
$d_{j,l}(a)$ behaves as a positive power of $a$,
\begin{equation}
 d_{j,l}(a)\sim \left(\frac{3}{l+1}\right)^{3/(2-2l)}
 (3a^2/\gamma\lP^2j)^{3(2-l)/(2-2l)} a^{-3}
\end{equation}
i.e.\
\begin{equation} \label{djasmall}
 d_j(a):=d_{j,\frac{3}{4}}(a)\sim \left(\frac{12}{7}\right)^6
 (\tfrac{1}{3} \gamma\lP^2 j)^{-15/2} a^{12}\,.
\end{equation}

This is the main ingredient for effective classical equations with the
most dramatic effects, and the methods developed here allow us to put
this term in effective equations on a more solid footing.  In order to
avoid numerical complications related to the additional degree of
freedom $\phi$ we use a dust model and employ the same effective
density for its energy density $\rho(a)=M/a^3$. We emphasize that for
dust this modification to the matter density is not the one most
naturally expected from loop quantum gravity. Rather, in loop quantum
gravity the matter {\em Hamiltonian} is primary and will be quantized.
Since this is a constant $M$ for dust, there would be no modifications
of this kind at all to the classical equations. We use the
modification (which can also be interpreted as arising from an
additional quantization ambiguity \cite{Golam,Robust}) here to model
the kinetic term of a scalar field Hamiltonian, and in order to study
the implications of effective densities.

\begin{figure}
 \includegraphics[width=14cm,keepaspectratio]{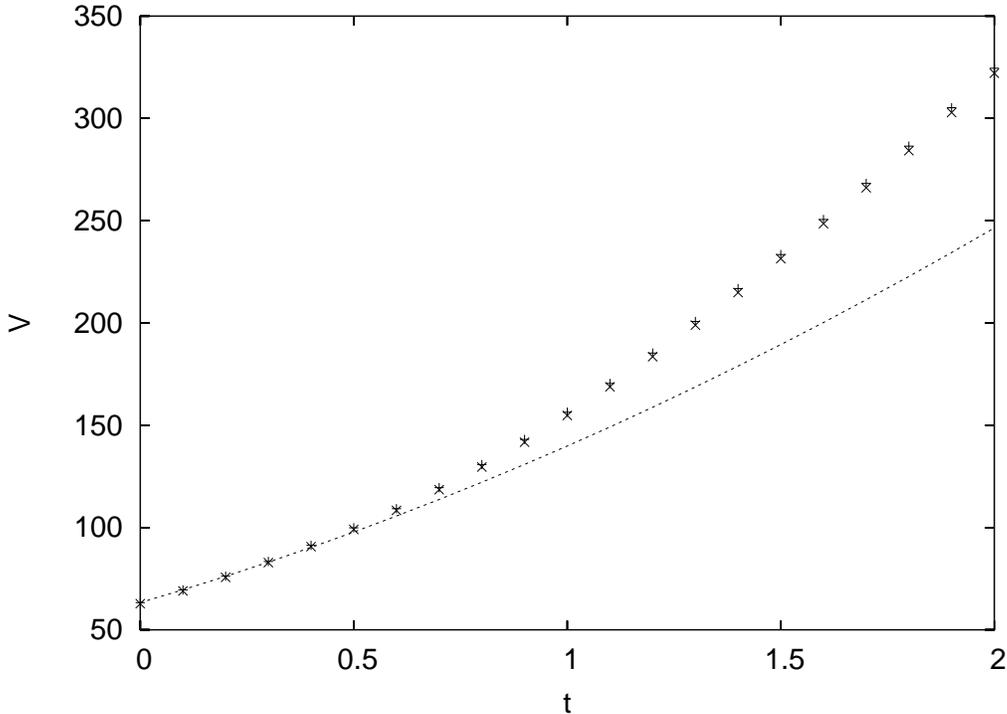}
 \caption{Expectation value of the volume ($+$) compared to the
 classical solution $V(t)$ (dashed) and the volume computed from the
 expectation value of the triad ($\times$) for dust with $M=10$ and an
 initial peak around $m_c=100$ ($N=1000$, $\sigma=20$). The ambiguity
 parameter $j$ for the effective density is $j=400$, such that the
 density peaks at $m_*\approx j/2=200$ corresponding to a volume
 $V_*=(2\gamma\lP^2 m_*/3)^{3/2}\approx 180 \lP^3$, which is reached
 around $t=1.1$. \label{Matter}}
\end{figure}

\begin{figure}[tbh!]
 \includegraphics[width=16cm,keepaspectratio]{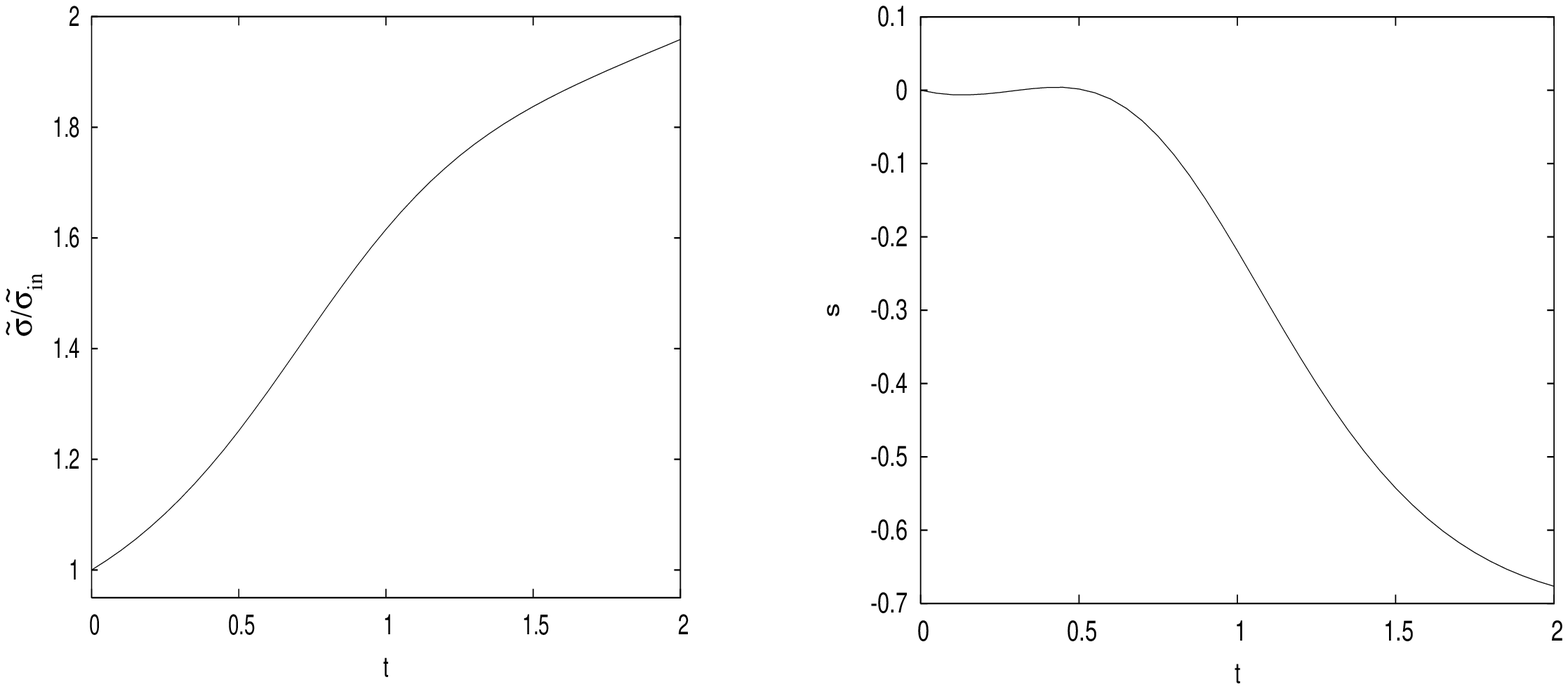}
 \caption{The evolution of standard deviation and skewness for the
   case in Fig. \ref{Matter}. The spread increases most strongly in
   the inflationary regime ($t<1.1)$), while the wave packet is skewed
   stronger during unaccelerated expansion.\label{Matter_spread}}
\end{figure}

\begin{figure}[tbh!]
 \includegraphics[width=16cm,keepaspectratio]{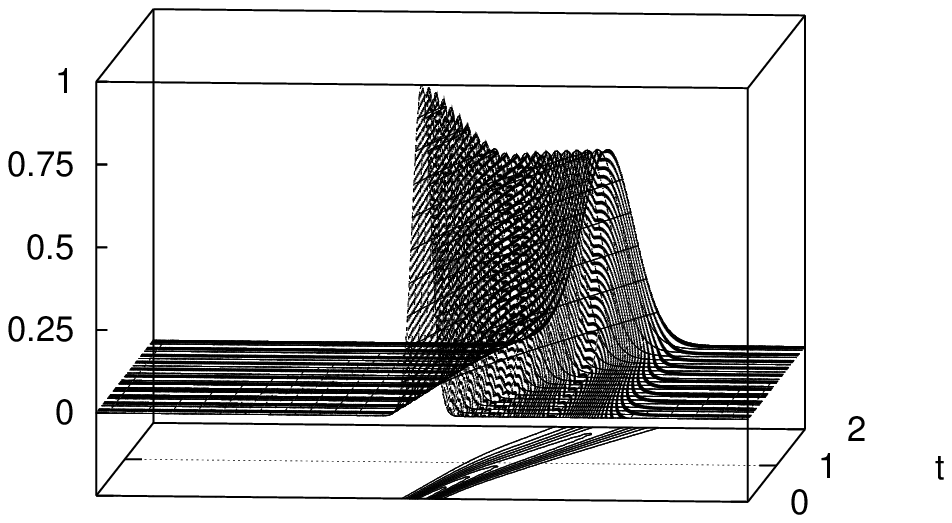}
 \caption{A spreading wave packet $|\psi_{t,i}|$ for the evolution shown in
Fig.~\ref{Matter}. \label{MatterWave}}
\end{figure}

To that end we replace the classical matter energy density for dust,
$\rho(a)=M/a^3$ with a constant $M$, by $\rho(a)=Md_{j,l}(a)$ and
study the evolution as determined by (\ref{diffHam}) and the
effective theory.  Note that unlike the case with a cosmological
constant, the effective theory is different from the classical one (in
which $\rho(a) = M/a^3$) due to the introduction of a modified geometrical
density. To see the differences between the two (which also highlights
the role of introducing $d_{j,l}$) let us first compare the volume
expectation values from (\ref{diffHam}) with the classical
theory. We have shown the results in Fig.~\ref{Matter}.  It is
interesting to see the way the effective density modifies the quantum
dynamics given by difference equations.
As is clear from plot, the classical theory does not match the quantum
description which indicates that the volume expectation value increases
more strongly than the classical solution, which unlike in
Fig.~\ref{LambdaSmall} is not an effect of a spreading wave packet or
skewness. Fig. \ref{Matter_spread} shows that spreading and skewness
of the wave packet are not large. This is also depicted in
Fig.~\ref{MatterWave} showing that wave packet does not spread
strongly during the displayed evolution.

From Fig.~\ref{Matter} it can be seen that the expectation value of
the volume operator agrees better with the cubic power of the
expectation value of the scale factor than with the classical
solution. This is another sign that the modified dynamics is
responsible for the deviations, rather than just change in shape of
wave packet during evolution. However, Fig.~\ref{Matter} does not tell
us decisively how well the expectation values would agree with
modified classical dynamics. A hallmark of the effective density
$d_j(a)$ is that it implies inflation
(accelerated expansion)
when inserted into the classical equations
and when the scale factor is below the peak value. In fact, the
expectation values in Fig.~\ref{Matter} increase more strongly than
the classical solution, but in the range $0<t<1.1$, where we are below
the peak, deviations are small.

\begin{figure}
 \includegraphics[width=14cm,keepaspectratio]{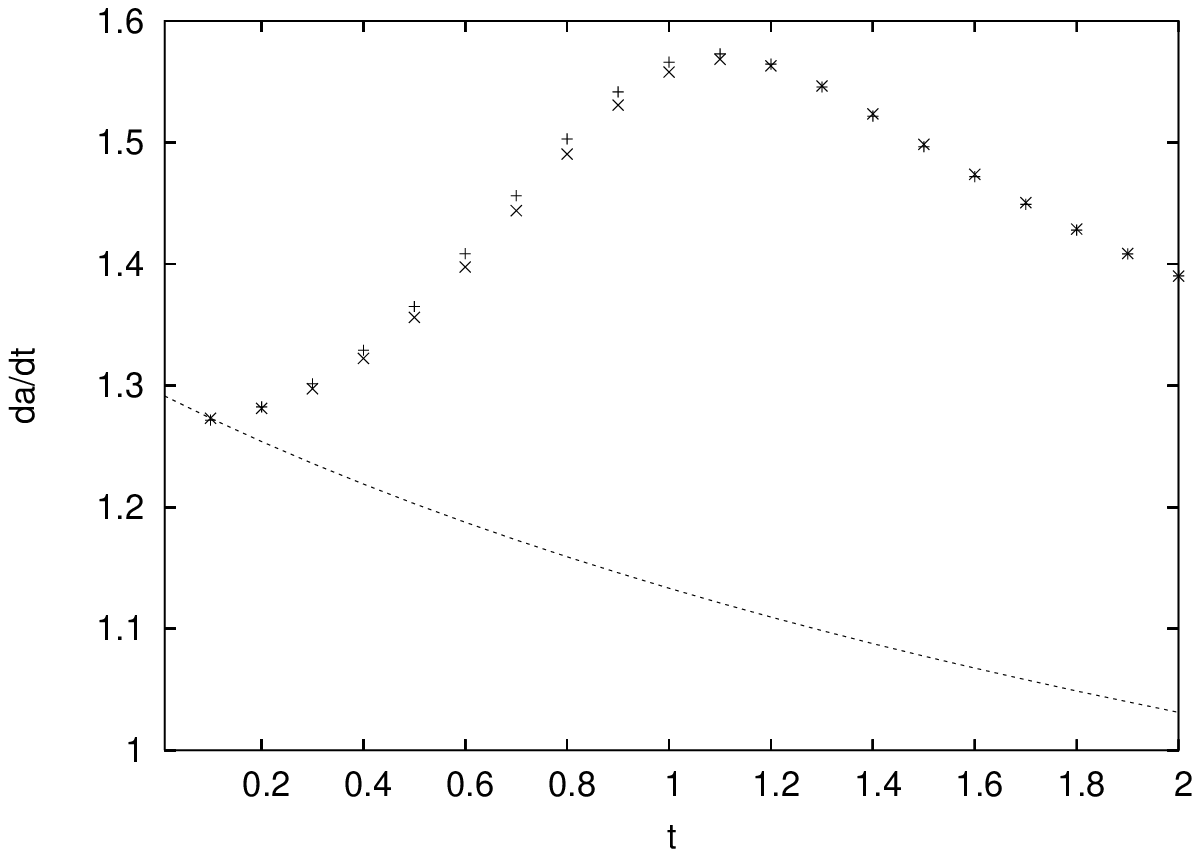}
 \caption{Time derivative of scale factors ($+$ corresponds to those computed from volume expectation
value and $\times$ refers to those of scale factor expectation values) plotted in
Fig.~\ref{Matter}. Dashed curve shows the classical curve. The classical description does not match the
evolution from quantum theory. \label{Matteradot}}
\end{figure}

It is much more illuminating to plot time derivatives, or rather
difference quotients from the numerically obtained data, and compare
with the classical behavior. Fig.~\ref{Matteradot} shows that in fact
the derivative of the scale factor increases when $t<1.1$, i.e.\ below
the peak of $d_j(a)$, and thus accelerates in agreement with the
expectation. This figure also shows that classical description completely
fails to capture the variation of scale factor with time as dictated by
quantum theory.

\begin{figure}
 \includegraphics[width=14cm,keepaspectratio]{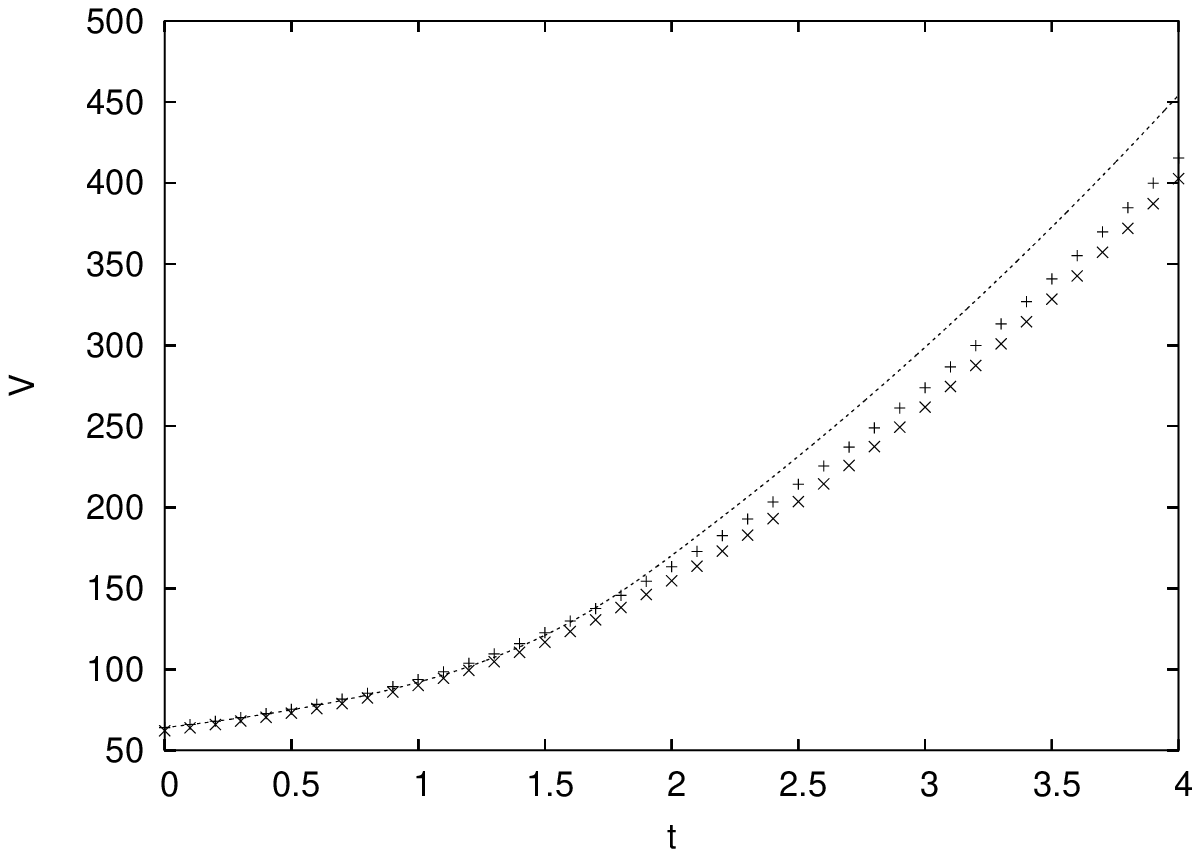}
 \caption{Volume expectation values ($+$ for $\langle \hat V\rangle$ and $\times$ for $\langle \hat a\rangle^3$)
 compared to the effective
   classical solution (dashed). Compared to the points in
   Fig.~\ref{Matter} the quantum behavior is different since the
   initial wave packet is now peaked on the effective rather than
   unmodified classical constraint surface (which decreases the
   initial $c_0$).  \label{Effective}}
\end{figure}

So far we have seen that the quantum evolution differs significantly
from the classical one and in particular leads to accelerating scale
factor expectation values and thus inflation. 
Now we turn our attention to the effective theory where $a^{-3}$ in the 
matter density is replaced by $d_j(a)$. We thus obtain a modified classical
description which as we will show agrees quite well with evolution determined
by difference equations.

\begin{figure}
 \includegraphics[width=14cm,keepaspectratio]{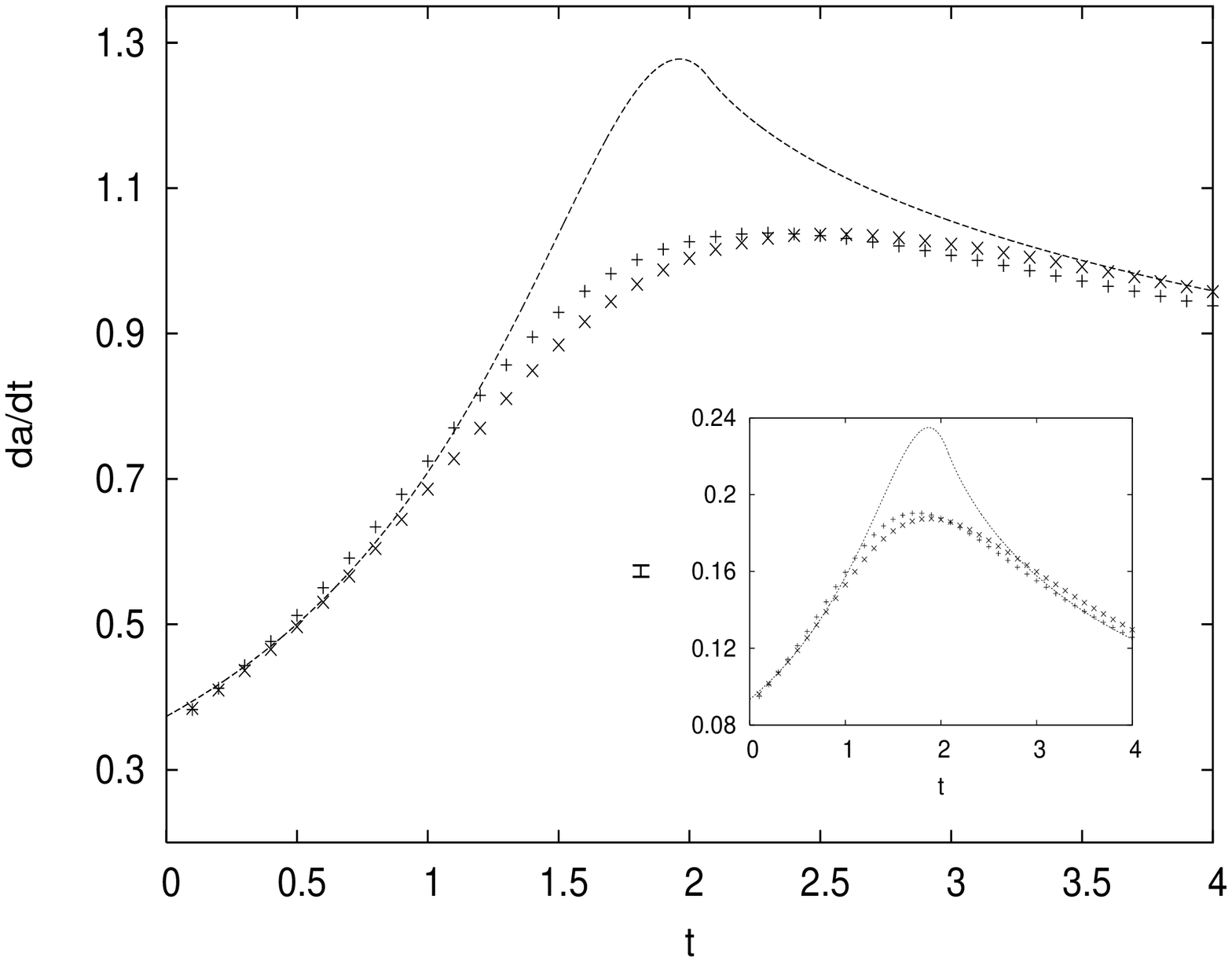}
 \caption{Time derivative of scale factors plotted in
 Fig.~\ref{Effective}. Inset shows variation of Hubble rate with time ($H = \dot a/a$). \label{Effectiveadot}}
\end{figure}

In order to check the effective density more directly we compare the
expectation values to numerical solutions of this effective classical
equation with the matter density replaced by $M d_j(a)$.
The result is shown in Fig.~\ref{Effective}. On comparison with
Fig.~\ref{Matter}, it is clear that the effective theory matches 
the quantum evolution much better than the classical description. 
We have plotted the derivative of scale factor with time in
Fig.~\ref{Effectiveadot}, which shows that time variation of scale factor
as computed from effective theory 
 agrees well with the change of the
expectation values of scale factor in the regimes before and after the peak. In
particular, the inflationary behavior in the modified region with
increasing $\dot{a}$ can be seen from the effective semiclassical as well
as the quantum solution. After the peak, both show the expected
non-inflationary behavior.

Interestingly, around the peak the effective classical time derivative
is larger than the change in expectation values, which is also the
reason why the effective classical volume is slightly larger than the
quantum volume at later times in Fig.~\ref{Effective}. At the point
where the peak occurs, $\dot{a}$ is largest so that higher order
corrections (higher powers of $\dot{a}$ in the Friedmann equation) are
expected to have the strongest influence. Those corrections have not
yet been included into effective semiclassical equations and we leave
a more detailed investigation of their effect around the peak for
future work. Nevertheless, one can see numerically that in the case
studied here the effect of higher order terms is negligible. This can
also be seen from the fact that those higher order terms arise as
powers in $c=-\frac{1}{2}\gamma\dot{a}$ which thanks to the smallness
of $\gamma$ remains sufficiently small compared to one throughout the
evolution. Moreover, for the flat model the Hamiltonian is invariant
under change of sign in $c$ such that the next higher order correction
is suppressed by a power $c^2$.

\begin{figure}
 \includegraphics[width=16cm,keepaspectratio]{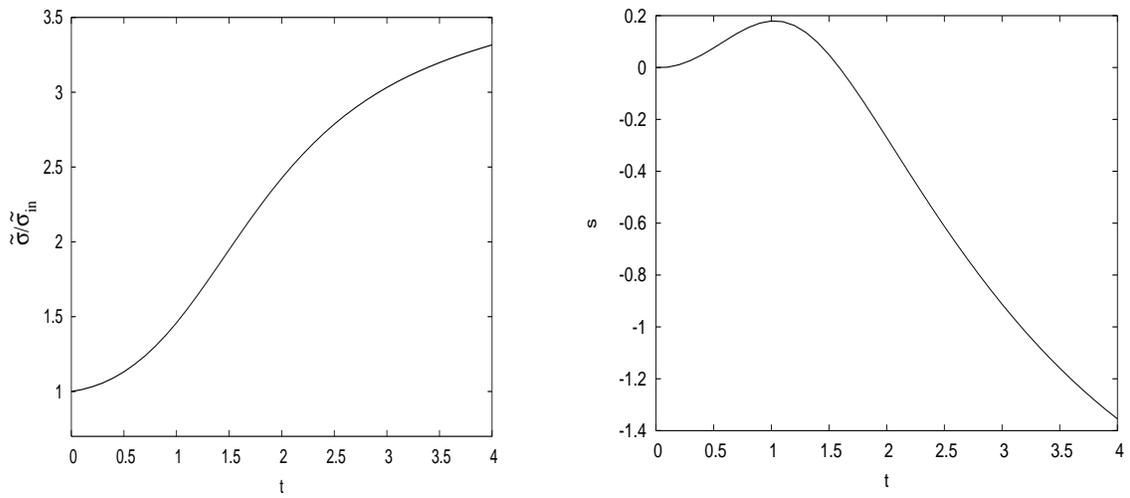}
\caption{The time variation of standard deviation and skewness for dust with effective density modification.  \label{Effective_spread}}
\end{figure}

Rather, the reduced derivative of the scale factor around the peak is
a consequence of deformations of the wave packet as shown in Fig.
\ref{Effective_spread}. In fact, around the peak the effective density
changes rather rapidly from increasing to decreasing behavior. Thus, a
part of the wave packet will already be in the decreasing regime while
its center is still in the increasing one, which lowers the overall
density seen by the wave. This behavior is verified by looking at the
skewness of the wave packet during the evolution which is plotted in
Fig.~\ref{Effective_spread}. First, the skewness turns positive which
means that the right tail of the wave packet becomes heavier than the
left one.  At some point, skewness starts to decrease and reaches
negative values, describing a redistribution of parts of the wave
packet such that now the left tail becomes more pronounced. This
redistribution implies that expectation values of powers of $m$, such
as the scale factor and volume, are lowered as compared to the
expected evolution. In fact, comparing Fig.~\ref{Effective_spread}
with Fig.~\ref{Effectiveadot} shows that the turn-around of skewness
starts just when the deviations between the change of
$\langle\hat{a}\rangle$ and the effective $\dot{a}$ appear, with the
expectation values increasing less strongly than the effective
classical value.


%

This observation shows that effective classical equations used so far
do not capture all details in the quantum evolution around the
peak. On the other hand, the behavior before and after the peak is
described very well. Cosmological studies so far have mainly focused
on the modified behavior at small scale factors and the initial
inflationary epoch, which is modeled reliably by using just the
effective density. The peak behavior was actually more problematic
since the Hubble parameter in the case of scalar dynamics easily
became dangerously high, larger than one in terms of Planck units,
which leads to doubt in the further semi-classical evolution. Here,
the quantum behavior with smaller $\dot{a}$ suggests that additional
effects from the wave packet can lead to a better semiclassical
picture, which may be modeled in effective classical equations by
taking into account effects of skewing wave packets. How this appears
in the case of a scalar field, and whether in this case $\dot{a}$ can
become large enough for higher order corrections to be relevant,
remains to be studied.

\subsubsection{Bounces}

\begin{figure}
 \includegraphics[width=14cm,keepaspectratio]{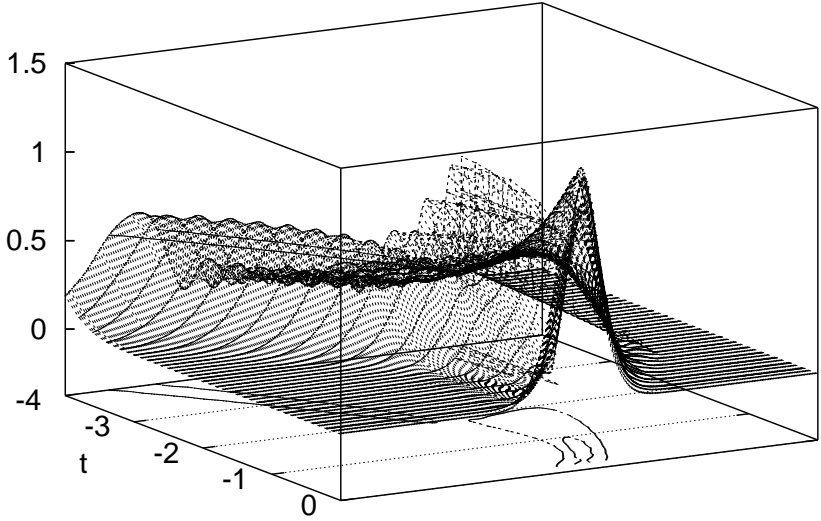}
 \caption{Wave packet evolving toward the classical singularity and
   bouncing off, only penetrating negligibly to negative $m$ (to the
   right). The parameters are $m_c=100$, $N=500$, $\sigma=10$, $M=10$,
   $j=200$. \label{BounceWave}}
\end{figure}

\begin{figure}
 \includegraphics[width=14cm,keepaspectratio]{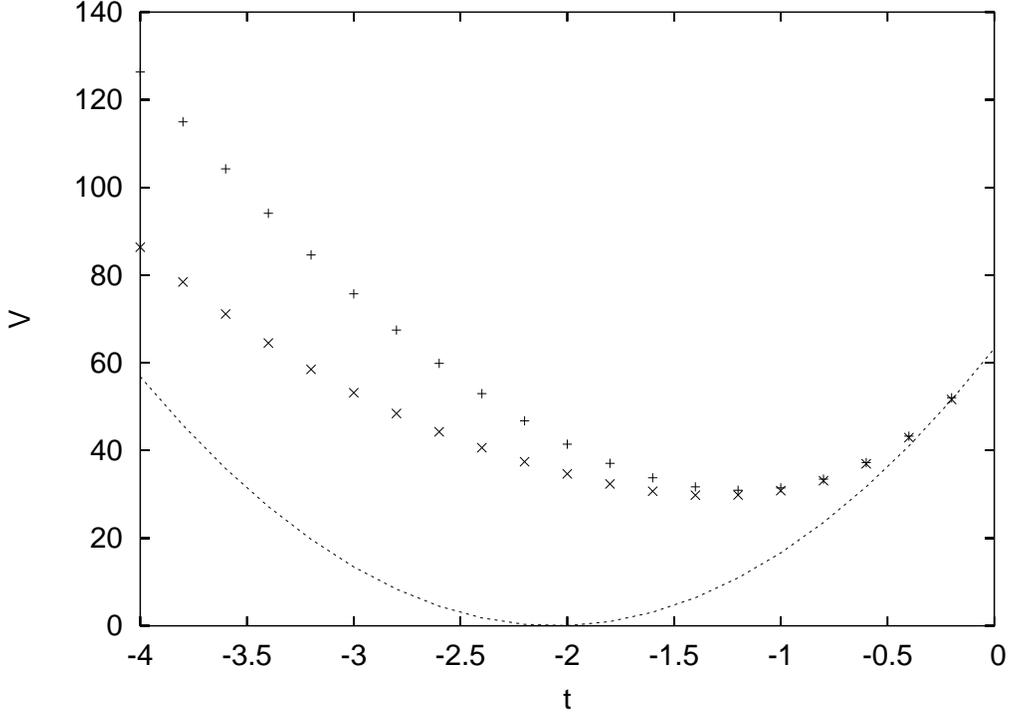}
 \caption{Volume expectation values for the wave packet in
 Fig.~\ref{BounceWave}. The expectation values bounce at non-zero
 values, while the classical curve hits zero
 volume. \label{Bounce}}
\end{figure}

So far we only looked at the evolution for rather large volume. For
smaller $|m|$, the approximation by classical evolution will become
worse and eventually break down. When exactly this is happening
depends on the parameters for the cosmological model and the choice of
initial wave function. As an example we again use the dust model but
evolve to earlier times, towards the classical singularity. In
Fig.~\ref{BounceWave} we show the evolution of the wave packet as it
deforms once a significant part of it reaches $m=0$. As in the case of
dust we first compare volume expectation values obtained from
(\ref{diffHam}) with the classical theory. The result is shown in
Fig.~\ref{Bounce}, which shows a bounce at non zero volume for
expectation values. The classical curve first hits the singularity at
zero volume.
%

\begin{figure}
 \includegraphics[width=14cm,keepaspectratio]{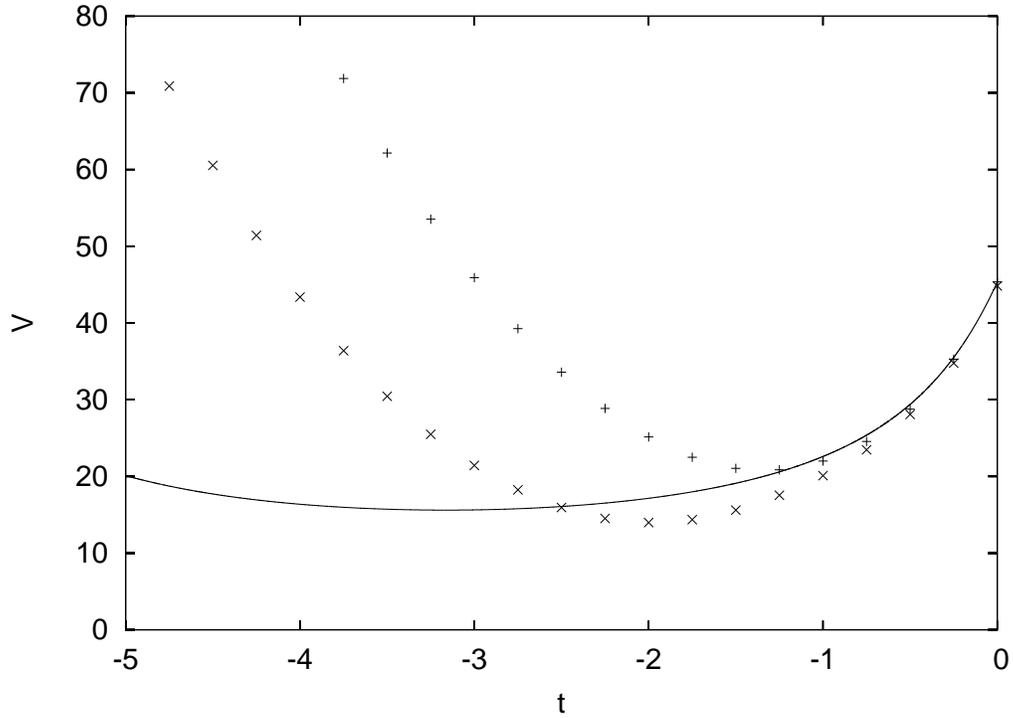}
 \caption{Volume expectation values for the wave packet initially
   peaked at the effective constraint surface compared to a numerical
   solution (solid line) of the effective Friedmann equation
   (\ref{FriedSigma}).  Since the corresponding wave packet, plotted
   in Fig.~\ref{BounceEffWave}, moves closer to the classical
   singularity than in Fig.~\ref{BounceWave}, the spread of wavefunction increases
   faster than in Fig.~\ref{Bounce}. \label{BounceEff}}
\end{figure}

It should be noted that the effective classical equations of a flat
model do not allow a bounce even when we use the effective density in
the dust case, and thus this bounce is not of the semiclassical type
as in
\cite{BounceClosed,Oscill,BounceQualitative,Cyclic,GenericBounce}. Further,
deformations of the wave function show directly that the classical
space-time picture is not valid during the bounce phase. Moreover, the
nature of a wave packet as opposed to sharp classical values becomes
more relevant, which provides a quantitative explanation of the
observed bounce. The spread $\sigma$ implies a correction to the
Friedmann equation which then takes the form
\cite{Perturb}
\begin{equation} \label{FriedSigma}
 \dot{a}^2+\tfrac{1}{4}\gamma^{-2}\sigma^{-2}=\tfrac{2}{3}a^2\rho(a)
\end{equation}
where for simplicity we  ignore the $a$-dependence of
$\sigma$ due to spreading.  Using the small-$a$
expansion
(\ref{djasmall}) for $d_j(a)$ in $\rho(a)=M d_j(a)$ and setting
$\dot{a}=0$ for the bounce results in a bouncing scale factor (in
Planck units)
\begin{equation} \label{abounce}
 a_{\rm bounce}\sim 3^{-25/28}\;(\tfrac{7}{4})^{3/7}\left(\tfrac{1}{8}
 \sigma^{-2} M^{-1} \gamma^{11/2} j^{15/2}\right)^{1/14}\approx
 2.44\,.
\end{equation}
Since we used the small-$a$
expansion,
which overestimates $d_j(a)$, the bounce value is slightly larger
resulting in $a_{\rm bounce}\approx 2.50$ if the function $d_j(a)$ is
used
in its full form.
The corresponding bounce volume, $V_{\rm
  bounce}\approx 15.6$ is considerably smaller than the minimum
expectation value in Fig.~\ref{Bounce}. But if we use the effective
Friedmann equation (\ref{FriedSigma}) to place the initial wave packet
on the effective constraint surface the expected bounce radius and the
numerical one in Fig.~\ref{BounceEff} agree within the limits of the
expectation values $\langle \hat V \rangle$ and $\langle \hat a
\rangle^3$. In fact, the wave packet bounces earlier than expected,
but around the expected volume. Thus, the spread-dependent correction
term explains why there is a bounce and gives a good
estimate
for the bounce radius.

\begin{figure}
 \includegraphics[width=16cm,keepaspectratio]{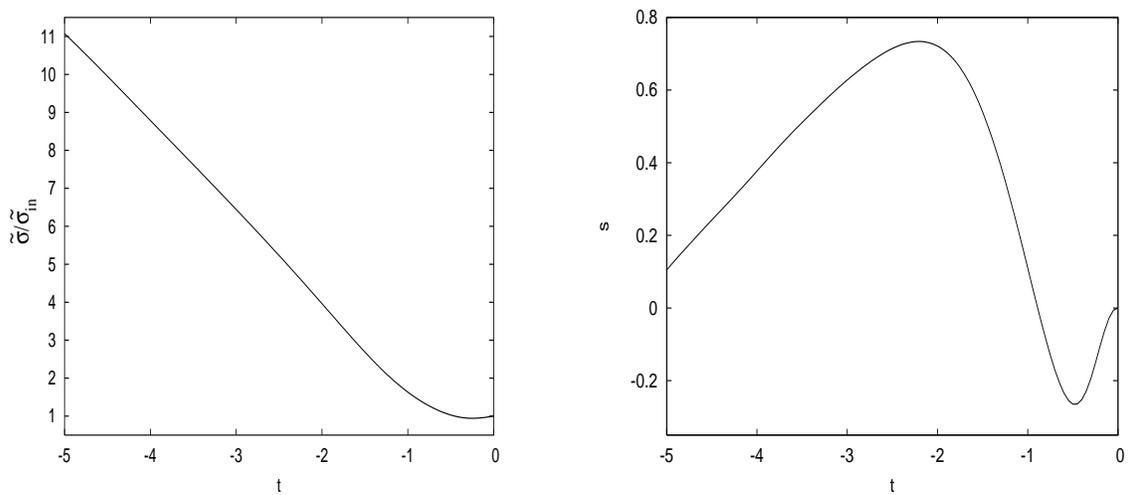}
 \caption{Increase in spread and skewness for backward evolution
 through a bounce as in Fig. ~\ref{BounceEff}.  \label{bounce_spread}}
\end{figure}

\begin{figure}
 \includegraphics[width=14cm,keepaspectratio]{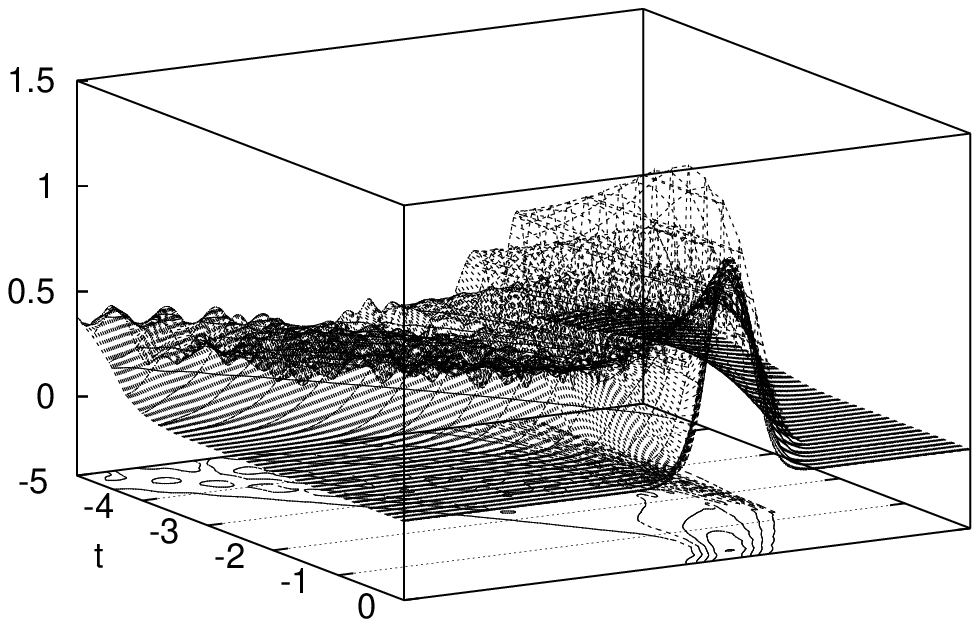}
 \caption{Wave packet evolving toward the classical singularity and
 bouncing off as in Fig.~\ref{BounceWave}, but initially peaked on the
 effective constraint surface at $m_c=80$. Since the bounce radius now
 is smaller, the leakage to negative $m$ is larger (see rightmost
 contour line). \label{BounceEffWave}}
\end{figure}

Nevertheless, it is also clear from Fig.~\ref{BounceEff} that the
agreement between the expectation values and the effective solution
deteriorates around the bounce. This can also be seen from
Fig. \ref{bounce_spread}, which shows considerable change in spread of
the wavefunction. In fact, the wave packet does not only spread but
also separates into different packets as seen in
Fig.~\ref{BounceEffWave}. Thus, even taking into account the spread
dependence in the modified Friedmann equation would not completely
describe the quantum behavior. (Note that the spread dependent
correction term was derived under the assumption of a Gaussian wave
packet. Fig.~\ref{bounce_spread} also shows that the skewness does not
increase strongly, but for this case of a wave separating into
different packets skewness alone is not a good measure for the
deviations from a Gaussian.) The evolution can then no longer be seen
as semiclassical and it is not possible to get a better agreement by
including further corrections into an effective equation.

\begin{figure}
 \includegraphics[width=14cm,keepaspectratio]{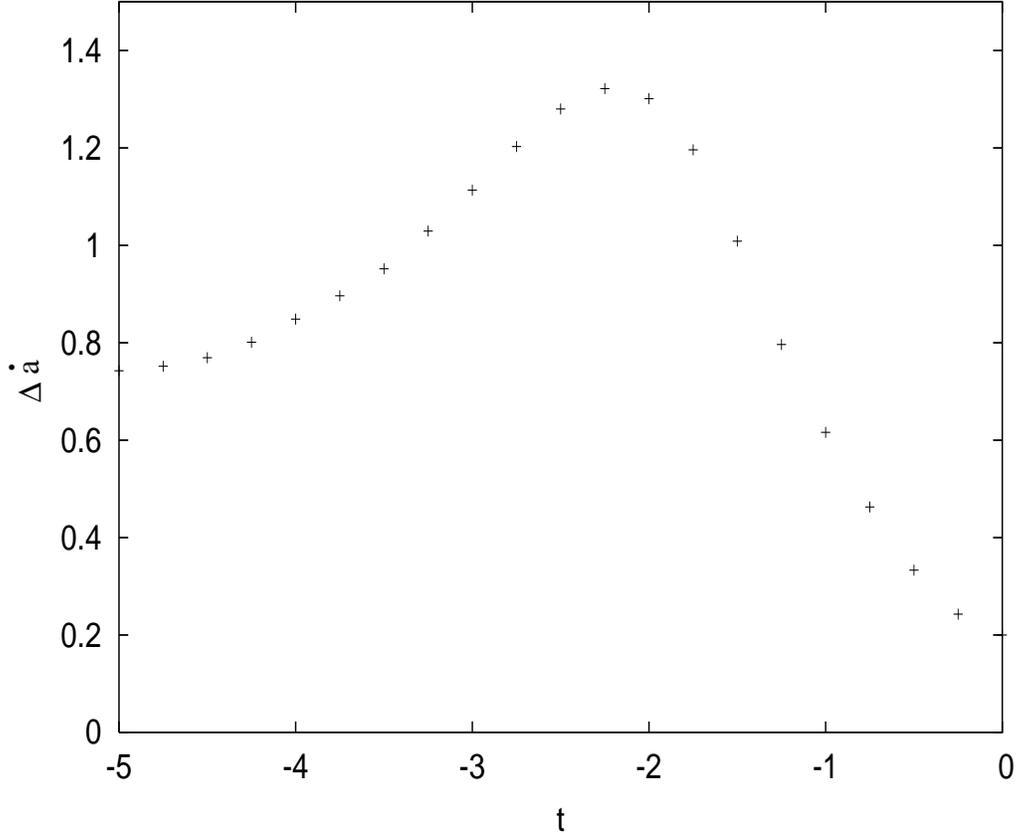}
 \caption{Extrinsic curvature fluctuation
 $\Delta\dot{a}=\sqrt{\langle(\hat{\dot{a}})^2\rangle-
 \langle\hat{\dot{a}}\rangle^2}$ corresponding to Fig.~\ref{BounceEff}
 where the operator for $\dot{a}$ is derived from the $c$-operator
 mapping a state $\psi_m$ to $\frac{1}{4}i
 (\psi_{m+1}-\psi_{m-1})$.  \label{Bouncecf}}
\end{figure}

\begin{figure}
 \includegraphics[width=14cm,keepaspectratio]{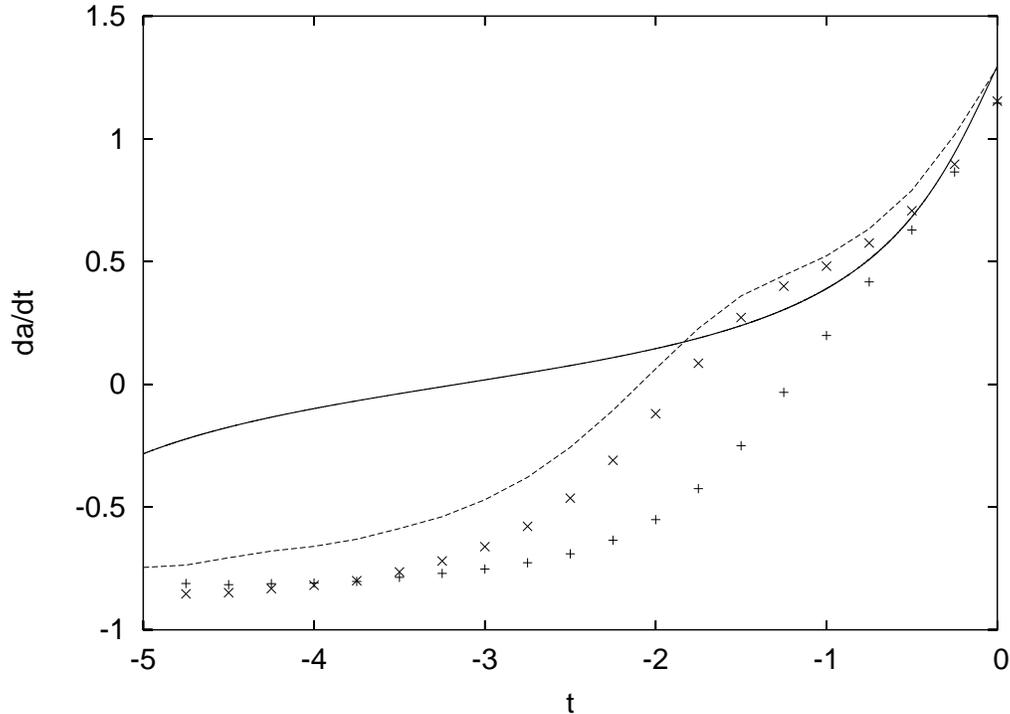}
 \caption{Time derivative of the scale factor expectation values in
   Fig.~\ref{BounceEff} compared to the effective solution (solid).
   The dashed line connects expectation values
   $\langle\hat{\dot{a}}\rangle$.
 \label{Bounceadot}}
\end{figure}

Still, one can also see that there is a rather undisturbed wave packet
bouncing off, while other parts of the wave function stay around the
classical singularity which does not affect expectation values of
geometrical quantities very much. Thus during the evolution shown here
the expectation value of the volume and the cube of that of the scale
factor do not deviate too much near the classical
singularity. However, since strong oscillations build up rapidly,
there are strong curvature fluctuations. This is shown in
Fig.~\ref{Bouncecf}, where the fluctuations of $\dot{a}$, computed
from the operator
\begin{equation} \label{adotop}
 (\hat{\dot{a}}\;\psi)_m = \tfrac{1}{2}i\gamma^{-1}
 (\psi_{m+1}-\psi_{m-1})\,,
\end{equation}
which initially are smaller than $\dot{a}$ increase to have a maximum
at the bounce. After the bounce, curvature fluctuations decrease but
stay larger than initially, and are comparable to the value of
$\dot{a}$ in Fig.~\ref{Bounceadot}.  We thus can still think of
semiclassical spatial slices of a certain volume at least in early
stages of the bounce, but since the extrinsic curvature is not sharp
one cannot think of them as forming a classical, smooth space-time.

If, as before, we compute the time derivative of the expectation value
of the scale factor, rather than the expectation value of curvature,
it is seen from Fig.~\ref{Bounceadot} that it is still increasing.
That is it accelerates, as a consequence of the effective density.
This figure also shows that the expectation value of the
$\dot{a}$-operator (\ref{adotop}) follows the change of $\langle
\hat{a}\rangle$ more closely than the effective classical solution,
which implies that the interpretation of extrinsic curvature (computed
from the geometrical quantity $a$ as compared to connection
components) has an unambiguous meaning.

As seen in Fig.~\ref{BounceEffWave}, the wave function leaks only
slightly into the part of minisuperspace corresponding to negative $m$
(i.e.\ the part corresponding to the side beyond the classical
singularity in the internal time formulation
\cite{Sing,IsoCosmo}). This is another reason for the fact that we can
still approximately speak of a classical bounce of the volume. If the
wave function at negative $m$ would be larger, also the spread in
spatial geometrical quantities and not just in extrinsic curvature
would be large and classical geometry would break down
completely. Whether or not this is happening depends very sensitively
on detailed quantization choices in the Hamiltonian constraint and
initial values of the wave function (see also
\cite{ScalarLorentz,BoundProp,ClosedExp}).

\section{Conclusions}

Splitting group averaging into two steps, first evolving an initial
state and then integrating along the group orbit, allows us to introduce
a coordinate time parameter into quantum gravity, although only in an
approximate sense. The resulting time-dependent family of states, the
state-time, does not solve the constraint exactly; only integrating
over time results in a physical state. Moreover, this does not
introduce a physical time parameter into the quantum theory, but only
coordinate time in semiclassical regime.
Still, the resulting evolution equations are
helpful in semiclassical analysis and in providing intuitive pictures
of quantum effects.

Our main application in this paper is a justification of using
effective densities as the most prominent quantum gravity effect
imported into effective classical equations. As it turns out, the
modified classical equations describe the quantum evolution of wave
packets very well, down to surprisingly small scales. Stronger
deviations become noticeable only when the wave packet starts to touch
the classical singularity. When exactly this happens depends on
details of the model and the chosen initial state. The diverse effects
giving rise to departures from classical behavior can be separated by
varying the parameters involved in the models. For instance, we chose
rather large values for the ambiguity parameter $j$ determining the
peak of the effective density, in order to distinguish this effect
from that of perturbative corrections. In this way it is possible to
study each correction term in the effective Hamiltonian separately.

The replacement of classically diverging factors of $a^{-3}$ in
dynamical equations by a bounded effective density $d_{j,l}(a)$ has been
confirmed as the main effect.  Effective densities themselves are
partly responsible for this observation since, via the effective
Friedmann equation, they lead to a reduction of extrinsic curvature
$\dot{a}$. This means that higher order corrections, i.e.\ higher
powers of $\dot{a}$, are less relevant at small $a$ than without
effective densities. Visible effects then occur most likely around the
peak of the effective density, where $\dot{a}$ is largest. As observed
here, the quantum Hubble rate is in fact smaller than would be
expected just from the effective density, which is helpful for
cosmological applications. However, numerically one can check that in
the cases studied here the reduced Hubble rate is a consequence of
deformations of the wave packet, parameterized here by the skewness,
rather than of higher order terms. Such an effect would have to be
included in effective classical equations by introducing a correction
term depending on the skewness of an evolving wave packet. That it is
possible to describe the influence of properties of wave packets on
the evolution by effective classical equations has been demonstrated
here by studying bounces implied by a spread dependent correction
term. The corresponding correction for skewness, however, is not known
since in the derivations of correction terms so far Gaussian states
with zero skewness have been used \cite{Perturb}. For cosmological
purposes it would be interesting to apply this technique to the case
of a scalar field and study the role played by effective densities in
particular around the peak.

Detailed formulas for perturbative correction terms, which include
higher order and higher derivative corrections, and uncertainty
correction terms related to the spread of the wave function, are
currently being evaluated. Once available, they can be used for a
direct comparison with the quantum evolution as studied here. In
particular the uncertainty corrections, which are proportional to
$\sigma^{-2}$ \cite{Perturb}, play a role at small volume since,
unlike higher order corrections, they are not suppressed by effective
densities. An application of this correction term can be found in
studying bounces, where it provides an explanation for the bounce in
Fig.~\ref{BounceWave} which does not follow from effective densities
alone. The expression (\ref{abounce}) of the bounce radius determines
the scale where quantum effects from the wave packet become important,
indicated by correction terms depending on the spread $\sigma$. For a
phenomenological analysis, $a_{\rm bounce}$ can be used as initial
value of the scale factor, which is relevant for estimates of the
amount of inflation. Since the value derived here for dust depends on
all the parameters of the model, in particular the ambiguity parameter
$j$, it is clear that the analysis will be more complicated than the
original ones, where the initial scale factor was assumed to be
$a_{\rm initial}=\sqrt{\gamma}\lP$. With an expression like $a_{\rm
  bounce}$, more reliable estimates can be made and constraints on the
parameters can be tightened. Thanks to the strong initial increase of
$d(a)$ with $a$, which is responsible for the small power $1/14$ in
(\ref{abounce}), the dependence on parameters of the model is,
fortunately, rather weak.

At very small scales the evolution of wave packets shows when the
classical space-time picture can be trusted, when it needs to be
corrected and when it breaks down completely. In this regime the
evolution becomes much more sensitive to quantization choices in the
Hamiltonian constraint. In particular this is true for the leakage
into the domain of negative $m$, where the orientation of space is
reversed, which is not surprising since this corresponds to evolution
beyond the classical singularity in the internal time picture. In this
context one should note that we had to use a symmetric ordering of the
constraint for the coordinate time evolution to be numerically stable.
This ordering already implies changes to the issue of initial
conditions \cite{DynIn,Essay} and on the relation between the wave
function at positive and at negative $m$. The alternative procedure of
using a lapse function $N(t)=a(t)$ and quantizing $NH$ symmetrically,
as mentioned in Sec.~\ref{NumImp}, also changes the issue of initial
conditions since the lapse vanishes at the classical singularity and
the wave function at negative $m$ completely decouples from that at
positive $m$. Thus, while the coordinate time picture is well suited
to justifying effective classical equations at non-vanishing volume,
the issue of the classical singularity can be understood only by using
the wave function directly and thus employing an internal time to
formulate evolution. Since the classical space-time picture breaks
down in this regime, there is no analog to coordinate time.

\acknowledgments We thank Kevin Vandersloot for useful discussions. PS
thanks Max-Planck-Institut f\"ur Gravitationsphysik for supporting a
visit and warm hospitality during early stage of this work.  His work
is supported in part by Eberly research funds of Penn State and by NSF
grant PHY-00-90091.

\bibliographystyle{../prsty}
\bibliography{../Bib/QuantGra}

\end{document}